\documentclass[twocolumn,prl,amsmath,amssymb,floatfix,superscriptaddress,showpacs]{revtex4} 
\usepackage{graphicx}% Include figure files
\usepackage{dcolumn}% Align table columns on decimal point
\usepackage{bm}% bold math

\begin{document} 

%\setlength{\topmargin}{0in}

\title{Magnetic field control of charge structures in the magnetically disordered phase of the multiferroic LuFe$_2$O$_4$}

\author{Jinsheng~Wen}
\affiliation{Condensed Matter Physics and Materials Sciences Department, Brookhaven National Laboratory, Upton, New York 11973}
\affiliation{Department of Materials Science, State University of New York, Stony Brook, New York 11794}
\author{Guangyong Xu}
\affiliation{Condensed Matter Physics and Materials Sciences Department, Brookhaven National Laboratory, Upton, New York 11973}
\author{Genda Gu}
\affiliation{Condensed Matter Physics and Materials Sciences Department, Brookhaven National Laboratory, Upton, New York 11973}
\author{S.~M.~Shapiro}
\affiliation{Condensed Matter Physics and Materials Sciences Department, Brookhaven National Laboratory, Upton, New York 11973}
\date{\today}

\begin{abstract} 
Using neutron diffraction, we have studied the magnetic field effect on 
charge structures in the charge-ordered multiferroic material LuFe$_2$O$_4$.
An external magnetic field is able to change the magnitude and 
correlation lengths of the charge valence order even before the 
magnetic order sets in. This affects the dielectric and 
ferroelectric properties of the material and induces a giant magneto-electric 
effect. Our results suggest that the magneto-electric coupling in LuFe$_2$O$_4$ 
is likely due to magnetic field effect on local spins, in 
clear contrast to the case in most other known multiferroic systems 
where the bulk magnetic order is important.
\end{abstract} 


\maketitle

Multiferroics are materials where magnetism and ferroelectricity, which are in many cases mutually 
exclusive, can coexist. In certain multiferroic systems, the situation becomes more appealing when 
ferroelectric polarizations appear to be connected with the magnetic order and can be manipulated 
using an external magnetic field~\cite{kimura}. This aspect of these materials has inspired tremendous 
interest because of the great potential for device applications~\cite{datastorage,Eerenstein,swcheong}.  
While the majority of these materials have spiral magnetic orders~\cite{kimura,lawes:087205,kimura:137201,
yamasaki:207204,taniguchi:097203}, normally resulting from geometric frustration, there are some 
exceptions with systems having collinear spin structures~\cite{PhysRevLett.93.177402,chapon:097601,
aliouane:020102,choi:047601}. However, they all share one common feature: electric dipole moments 
are always induced by the formation of inversion-symmetry-breaking magnetic order, where an external 
magnetic field can affect the magnetic structure and therefore change the ferroelectric properties. 
There is another type of multiferroic where ferroelectricity and magnetism develop more independently, 
such as BiFeO$_3$ and BiMnO$_3$. In these systems the ferroelectricity mainly comes from the shifts of the 
Bi ions while magnetism is a result of Fe/Mn moments. At low temperature the two orders coexist with 
reasonably large electric polarization and magnetization, but the magneto-electric response is very weak.

\begin{figure}[ht]
\includegraphics[width=\linewidth]{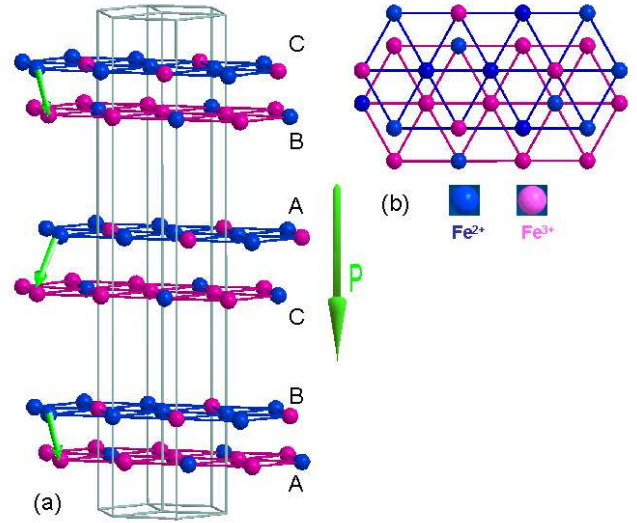} 
\caption{(Color Online) Structure of the multiferroic LuFe$_2$O$_4$, showing the Fe atoms only. (a) The three double layers. 
In the charge-ordered phase, because of the charge imbalance (e.g. here the top and bottom layers 
each has Fe$^{2+}$:Fe$^{3+}$ ratios of 2:1 and 1:2), each double layer gives rise to local electric 
dipole moments. With an external electric field, a bulk ferroelectric 
polarization {\bf P} can be induced. 
(b) A top view of the charge order in on Fe-O double layer. This arrangement have a periodicity of 3 
along the [110] direction, and gives rise to the (H/3, H/3, L/2) type supper-lattice charge peaks.}
\label{fig:1}
\end{figure}

LuFe$_2$O$_4$ is a new multiferroic material where the ferroelectric polarization originates from valence 
order of Fe$^{2+}$ and Fe$^{3+}$ ions instead of cation displacements as in conventional ferroelectrics. 
The ferroelectricity starts to appear slightly above room temperature where no magnetic order is present 
in the system~\cite{ikeda}. As far as the multiferroicity is concerned, 
LuFe$_2$O$_4$ belongs to neither 
of the above two categories. Here Fe$^{2+}$ and Fe$^{3+}$ ions form double 
layers (Fig.~\ref{fig:1}a), and the non-uniform 
charge structure contributes to local electrical polarizations. 
The bulk ferroelectric polarization {\bf P} 
appears first when the system enters a three-dimensional (3-D) 
charge-ordered phase~\cite{ikeda} at the 
charge-ordering temperature T$_{CO}$ around 340~K (Fig.~\ref{fig:1}b). 
Although the polarization is not induced 
by magnetic order, a significant change in {\bf P} is still observed when 
the system enters a long-range 
ferrimagnetic ordered phase~\cite{ikeda} at the magnetic ordering 
temperature T$_N$ around 240 K, suggesting a 
large coupling between the ferroelectric polarization and the ordering 
of Fe spins. From the perspective of making 
practical devices, LuFe$_2$O$_4$ is ideal since it has large dielectric 
response and magneto-electric 
coupling~\cite{Subramanian} at room temperature. In this letter, 
we present neutron scattering work on the 
charge order in the system. Our findings show that the static charge 
structure in LuFe$_2$O$_4$ can be 
affected by the application of an external magnetic field in the 
non-magnetic phase, which is extremely unusual. Our results naturally explain 
the unprecedented giant magneto-electric response at room temperature. 
Furthermore, the fact that the magnetic 
field can directly affect the charge structure without any intermediate 
(magnetic) order suggests that a different
magneto-electric coupling mechanism has to be considered for 
this charge-ordered multiferroic system.

\begin{figure}[ht]
\includegraphics[width=\linewidth]{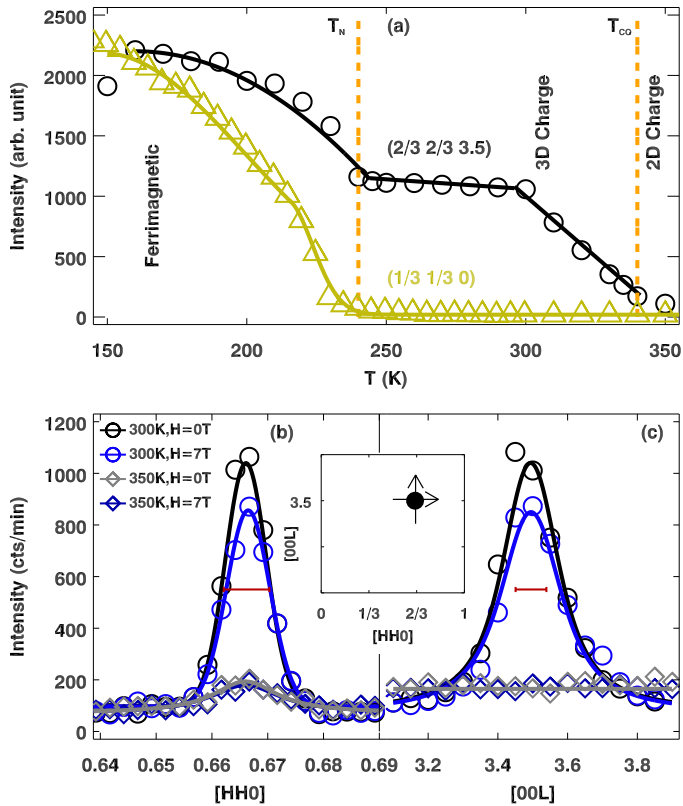} 
\caption{(Color Online) Magnetic and charge order in LuFe$_2$O$_4$. (a) The magnetic and charge Bragg peak intensities as order 
parameters vs. temperature. (b) and (c) Linear intensity profiles of the charge super-lattice peak (2/3, 2/3, 3.5) 
along [110] and [001] directions, respectively. The horizontal bars indicate the instrument resolution along the 
measured directions. The inset denotes the direction of the scans performed.}
\label{fig:2}
\end{figure}

Single crystals of LuFe$_2$O$_4$ are grown using the floating zone 
technique~\cite{crystalgrowth}. The typical 
crystal sizes are $\sim10\times5\times3$~mm$^3$. 
Our neutron scattering measurements 
are performed on BT9 triple-axis-spectrometer at the NIST Center for 
Neutron Research (NCNR). An incident neutron energy of 14.7~meV was selected 
by a pyrographic (PG002) monochromator, 
with beam collimations of 40’-40’-40’-80’, and another PG002 crystal 
used as the analyzer. PG filters are also 
used before the sample to reduce background from higher order neutrons. 
LuFe$_2$O$_4$ has a hexagonal structure and 
the sample has been oriented so that 
the horizontal diffraction plane is the (HHL) plane, 
which is defined by the vectors [110] and [001]. 
A magnetic field can be applied along the vertical $[1\bar{1}0]$ direction. 

The magnetic Bragg peaks in this compound can be measured at 
reciprocal space positions such as (1/3,1/3,L) and (2/3,2/3,L) 
for both half integer L and integer L 
values~\cite{twodspin,christianson:107601}, while the charge 
peaks only appear at half-integer L values. In Fig.~\ref{fig:2}a, 
we plot the magnetic order parameter [intensity measured at (1/3,1/3,0)],  
together with the charge order parameter [intensity measured at (2/3,2/3,3.5)].
Here we show that with cooling, the intensity of the
magnetic Bragg peak (1/3, 1/3, 0) only starts to rise around T$_N$ $\sim$ 240~K,
in good agreement with previous reports~\cite{twodspin,christianson:107601}.

Charge order in this system starts to appear at much higher temperature, when 
Fe$^{2+}$ and Fe$^{3+}$ are believed to form the structure shown in 
Fig.~\ref{fig:1}b~\cite{PhysRevB.62.12167,zhang:247602,angst-2008}. 
Super-lattice peaks 
arising from modulations of local atomic distortion due to the charge 
order can be detected near half integer 
L values, as shown by both x-ray and electron diffraction 
measurements~\cite{PhysRevB.62.12167,
zhang:247602,angst-2008} which are only sensitive to lattice 
structures and not affected by magnetic moments. 

We choose to monitor the supper-lattice peak 
(2/3,2/3,3.5) for the charge valence order. At temperatures 
above T$_N$ the situation is relatively simple since there is 
no magnetic contribution to the intensity.
The intensity profile for the charge super-lattice peak around 
(2/3,2/3,3.5) at 350~K
is shown in Figs.~\ref{fig:2}b and \ref{fig:2}c, with a peak 
along the [HH0] direction in the hexagonal plane. 
The scattering intensity remains constant when we scan along 
the out-of-plane direction [00L] near L=3.5, 
suggesting a two-dimensional (2-D) nature of the ordering 
at this temperature. On cooling below 340~K, the 
system enters a three-dimensional (3-D) charge-ordered phase, 
evidenced by the peak in the L-scan across the 
charge super-lattice peak. When these local electric dipoles 
arising from charge imbalance of Fe-O double layers 
order, spontaneous polarization starts to form and bulk 
ferroelectricity can be induced~\cite{ikeda}. The charge 
peak intensity appears to become saturated below 300~K. 
The sudden intensity increase at T$_N$ corresponds to superimposed 
magnetic scattering intensity at this wave-vector due to the 
magnetic order. 

\begin{figure}[ht]
\includegraphics[width=\linewidth]{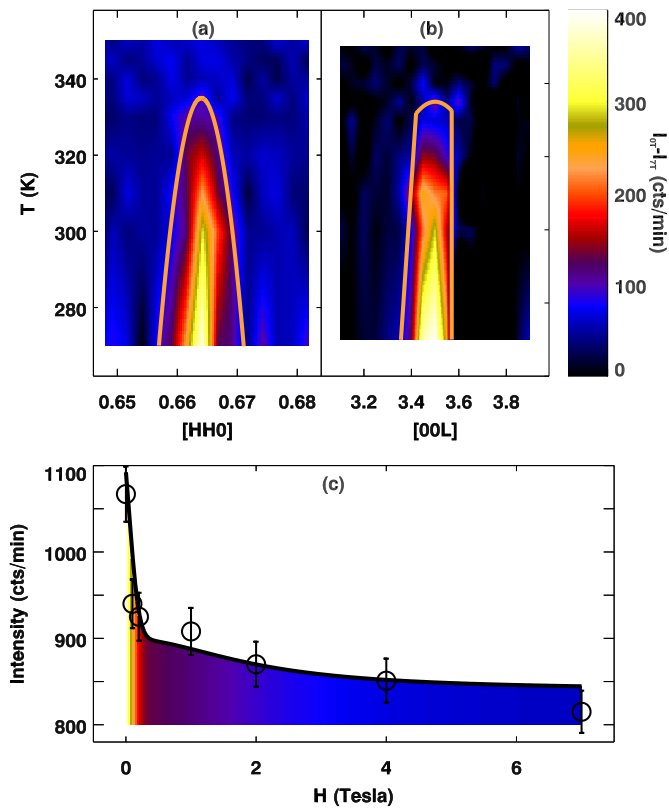} 
\caption{(Color Online) Magnetic field response on the charge super-lattice peak (2/3, 2/3, 3.5). 
(a)  and (b) Temperature dependence of the intensity difference between measured linear 
intensity profiles across the charge peak, performed with ZFC and FC with H=7~T.  
(c) Charge peak intensity measured at 300~K with FC from 350~K at different magnetic field strengths.}
\label{fig:3}
\end{figure}

To investigate the response of the system to an 
external magnetic field, we have repeated the measurements 
under a magnetic field cooling (FC) condition. A magnetic field 
of H=7~T has been applied at 350~K, along the 
$[1\bar{1}0]$ direction, perpendicular to the scattering plane. 
Here in the 2-D charge-ordered phase, no impact 
of the field has been observed (see Figs.~\ref{fig:2}b and \ref{fig:2}c) 
compared to the measurements under 
zero-field-cooling (ZFC). With further cooling under field, a partial 
reduction of the charge peak intensity is 
clearly evident. At 300~K, the charge peak intensity is reduced by 
about 25\% with FC (H=7~T), as shown in 
Figs.~\ref{fig:2}b and \ref{fig:2}c. The temperature dependence of 
the magnetic field effect is given in 
Figs.~\ref{fig:3}a and \ref{fig:3}b, where the intensity differences 
between the ZFC and FC (H=7~T) 
measurements have been plotted as a function of temperature. The field 
effect starts to be visible around 340~K 
when the system enters the 3-D charge-ordered phase, and continues 
to grow with further cooling. These results suggest a direct magnetic field
effect on the charge valence order since there is no evidence of 
field induced magnetic order above T$_N$. 
No magnetic peak intensity is present at (1/3, 1/3, 0) 
with FC of H=7~T in this 3-D charge-ordered phase; and the magnetic phase 
transition temperature remains the same as 
indicated by magnetization measurements~\cite{wang-2007,wen:unp}.

\begin{figure}[ht]
\includegraphics[width=\linewidth]{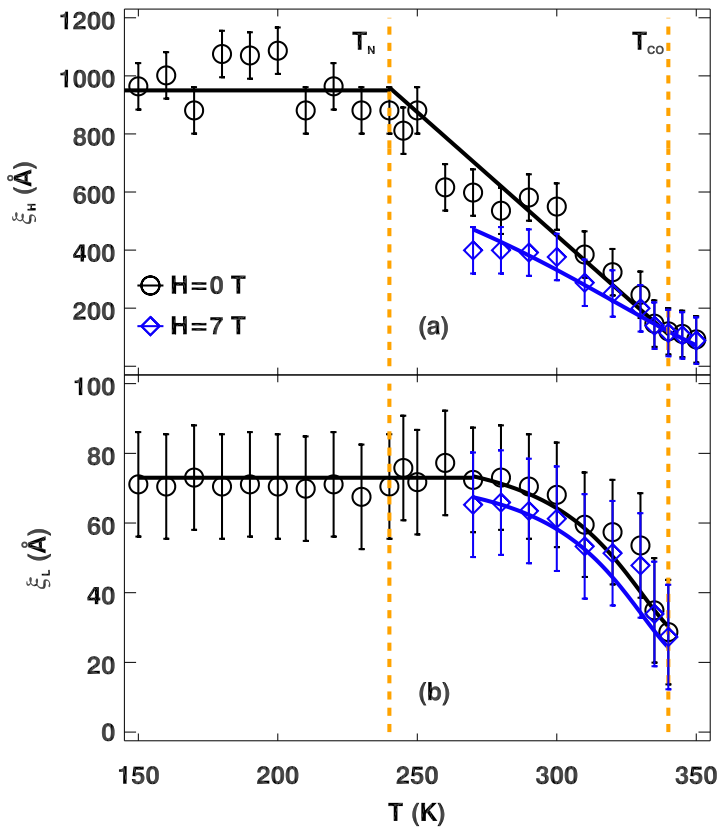} 
\caption{(Color Online) Charge order correlation lengths derived from the charge peak (2/3, 2/3, 3.5). 
(a) The correlation length in the hexagonal plane along [110] direction. (b) The out-of-plane 
correlation length along [001] direction. The correlation length is obtained as $\xi=1/\Gamma$, 
where $\Gamma$ is the half-width at half maximum of the Lorentzian function, which was used to 
(after convolution with the instrument resolution function) fit the linear intensity profiles 
of the charge super-lattice peak. Note that magnetic intensity starts to appear at this 
wave-vector when the system orders at T$_N$ around 240~K, and therefore the numbers below 
the magnetic ordering temperature is only given as a reference and should not be taken as 
having the same physical origin (the correlation length for the charge order) as those 
in the 3-D charge-ordered phase above T$_N$.}
\label{fig:4}
\end{figure}

The magnetic field effect on the charge order is strongly history dependent. 
Applying a field of H=7~T directly at 300~K 
after a ZFC process does not have any measurable effect on the charge peak 
intensity; neither does removing the field at 300~K 
after a FC process with H=7~T restore the charge peak intensity (back to 
the ZFC value). The external field can only affect 
the charge structure when applied above the 3-D charge ordering 
temperature before a FC process. In Fig.~\ref{fig:3}c, the charge 
peak intensity at 300~K is plotted for FC processes with different 
magnetic field strengths (field applied at 350~K). For magnetic 
field as small as H=0.1~T, a significant ($\sim$15\%) reduction on 
the charge peak intensity is already present. 
In addition to the intensity reduction, a change in the range over 
which the charge order can be maintained is also observed in 
the FC process. In Figs.~\ref{fig:4}a and \ref{fig:4}b, we show 
the instrumental resolution corrected in-plane and out-of-plane 
correlation lengths $\xi_H$ and $\xi_L$ of the charge order vs. 
temperature.  In the 3-D charge-ordered phase, the in-plane 
order appears to be long-range with $\xi_H$ approaching 800~\AA, 
while the order remains short-range along the out-of-plane direction L, 
with $\xi_L$ around 70~\AA. When cooling in a 7~T magnetic field, both 
$\xi_H$ and $\xi_L$ are affected. The magnetic field 
reduction of the charge order correlation lengths starts at 
T$_{CO} \sim$ 340~K and becomes more prominent with cooling.

What we have observed in this non-magnetic charge-ordered phase of 
LuFe$_2$O$_4$ is extremely unusual. As (static) charges do not 
respond directly to an external magnetic field, one generally does 
not expect charge order to be affected by a magnetic field. 
Of course, there are some exceptions. For instance, charge order can 
coexist with ferromagnetic order in manganites with colossal magnetoresistance 
(CMR)~\cite{PhysRevLett.81.3972,uehara,Loudon}. An external magnetic 
field can affect the charge order indirectly by modifying the 
magnetic order in the system~\cite{uehara}. Another example involves charge/spin stripe order in cuprate high temperature superconductors. 
Magnetic field dependence of charge stripe order in La$_{2-x}$Ba$_x$CuO$_4$ (LBCO, $x \approx 1/8$) has been reported~\cite{kim} at T = 2~K, 
but was attributed to the magnetic field suppression of superconductivity. In other words, in the few rare cases where charge orders 
can be affected by an external magnetic field, the effect is always mediated by another order (magnetic/superconducting) being 
modified by the (magnetic) field. In LuFe$_2$O$_4$, no such intermediate order is present between T$_N$ and T$_{CO}$, and the magnetic 
field appears to directly affect the charge structure. 

This has deep implications for magnetic field control of ferroelectric properties in multiferroic materials. 
The reported giant magneto-capacitance effect, where the dielectric response can be reduced by as much as 25\% 
by external magnetic field at room temperature~\cite{Subramanian} for LuFe$_2$O$_4$, had been originally attributed 
to the field affecting charge fluctuations. Our results suggest that the room temperature magneto-electric effect 
in this multiferroic system is a direct result of the field affecting the static charge order. When charge order 
is reduced, the magnitude of the resulting local electric dipole moments becomes smaller, and therefore the dielectric 
response is smaller. The fact that this giant coupling occurs at room temperature, in a non-magnetic phase, in clear 
contrast to most other magneto-electric coupling effects in known multiferroic systems where a magnetic order needs to 
be present to mediate the coupling, makes it extremely appealing and important. 

Although there has been some theoretical consideration~\cite{nagano:217202,naka:224441,xiang:246403} for the magnetic field 
effect on charge structure in LuFe$_2$O$_4$, most of these efforts are focused on the low temperature 
structure where the system orders magnetically. One possibility has been raised in these work that the Zeeman splitting 
of different spin states caused by an external magnetic field could affect the stability of various charge structures 
and therefore the dielectric response. In the low temperature magnetically ordered phase, a clear coupling between the 
charge and magnetic order has been observed~\cite{angst-2008} in LuFe$_2$O$_4$. At room temperature there is no long 
range magnetic order, yet it is possible that Fe ions could form local spin clusters which favor different charge 
order states depending on the spin structure of these clusters. We speculate that an external magnetic field can 
affect these local spin structures, and destabilize the ordered charge structure shown in Fig.~\ref{fig:1}b 
(or even stabilize new charge structures), and therefore affect the overall ferroelectric properties of the material. 
It is truly surprising though that interactions on a local scale can result in such a giant (magneto-electric) effect. 
The results reported here provide a grand challenge for theories to explain this new mechanism of magneto-electric 
interaction which can have significant impact on both the understanding of electronic structures in a magnetically 
disordered environment as well as development of new valence driven functional multiferroic materials.

\begin{acknowledgements} 
We thank W. Ratcliff, C. L. Broholm, and J. M. Tranquada for discussions. 
Work at Brookhaven National Laboratory is supported by U.S. Department of Energy Contract DE-AC02-98CH20886. 
\end{acknowledgements}

%\bibliography{lfo}

\begin{thebibliography}{29}
\bibitem[{wen()}]{wen:unp}
\bibinfo{note}{The situation becomes more complicated when the system orders
  magnetically at T$_N$ - the scattering intensity at (2/3,2/3,3.5) now has
  contributions from both magnetic order and lattice modulation due to the
  charge order. Our preliminary work~\cite{wen:unp} shows that below T$_N$, the
  magnetic structure is also affected by the external field, with partial
  suppression of magnetic intensities at integer L values and enhancement of
  magnetic intensities at half integer Ls. Detailed study is clearly required
  in the future to understand the complicated low temperature magnetic and
  charge structure under field, and we will only focus on the field effect in
  the 3-D charge-ordered phase above T$_N$ in this letter.}
\end{thebibliography}

\end{document}